\documentclass[onecolumn]{revtex4}
\usepackage{amssymb,amscd,amsmath,graphicx}
\begin{document}

\title{On the two principal curvatures as possible entropic barriers in a mesoscopic nonequilibrium thermodynamics model of complex matter agglomeration}  
\author{A. Gadomski$^a$}
\author{J.~M. Rub\'{\i}$^b$}  
\affiliation{
$^a$Institute of Mathematics and Physics, 
  University of Technology \\ and Agriculture,
85796 Bydgoszcz, Poland\\    
$^b$Departament de F\'{\i}sica Fonamental, Universitat de Barcelona, \\
08028 Barcelona, Spain  
}
%


\begin{abstract}
Matter agglomeration mesoscopic phenomena of irreversible type are well described by  nonequilibrium thermodynamics formalism. The description assumes that the thermodynamic (internal) state variables are in local equilibrium, and uses the well known flux--force relations, with the Onsager coefficients  involved, ending eventually up at a local conservation law of Fokker--Planck type. 
One of central problems arising when applying it to the matter agglomeration phenomena, quite generally termed the nucleation--and--growth processes, appears to be some physically accepted identification of entropic barriers, or factors impeding growth. 
In this paper, we wish to propose certain geometric--kinetic obstacles 
as serious candidates for the 
so--called entropic barriers. Within the framework of the thermodynamic formalism offered they are always 
associated with a suitable choice of a physical potential governing the system. 
It turns out that a certain choice of the potential of Coulomb (or, gravitational) type leads to emphasizing the role of the Gaussian curvature while another choice in a form of  the logarithmic physical potential results unavoidably in a pronounced role of the mean curvature. The whole reasoning has been tested successfully on a statistical--mechanical polycrystalline evolution model introduced some years ago for physical--metalurgical purposes, and modified for a use in biophysical soft--matter agglomerations. 
\end{abstract}


\maketitle

\section{Introduction}

In this work, we wish to focus on a matter agglomeration process ranging in size of a 
constituting entity (grain; crystallite; cluster) between micrometer (physical metallurgical 
specimen or colloidal flock) and nanometer (biophysical agglomerate {\it viz} biomembrane) scales. The systems that we are going to address possess the following characteristic basic properties \cite{shvind}
\begin{itemize}
\item They are diffusion--migration systems, which means that they are well--characterized by a diffusion process along the axis of their internal variable (volume, size, alike), and they undergo a kind of migration being recognized as a drift by capillary forces, provided that they are vitally present in the system \cite{hweaire}; 
\item The growing processes that are 'in action', are of irreversible nature, so that a thermodynamic formalism is possible to apply \cite{mweaire}; 
\item The diffusion coefficient is proportional to the surface magnitude of a basic entity constituting the agglomerate under evolution, a physical scenario well--known mostly in colloidal agglomerations \cite{novikov}; 
\item The drift term is always proportional to the curvature: It turns out that it is either very related with the mean curvature, what is supposed to be true in physical--metallurgical systems being of micrometer size of its basic entity, or it appears to be identified with the Gaussian curvature, which can reasonably be judged as being characteristic of nanometer scale rather, in which (bio)polymeric agglomerations are supposed to be in favor, 
look at Fig. 1; 
\item The systems we examine always obey the local conservation  law; 
\item They are constrained systems, mostly due to: (i) boundary conditons applied: they can be either of normal ($0$-Dirichlet) or of abnormal types \cite{mullins}; (ii) their total (hyper)volumes\footnote{Accept that the description considered is a space dimension ($d$)--dependent approach.} which they can preserve (surface tension or elasticity effects pronounced) or cannot do it (no special role of surface tension assumed) \cite{liu}. (Their evolutions, however, do not depend essentially upon the initial condition chosen.)
\end{itemize}
 
As the so-called case or exemplified study here, we wish to choose a physical--metallurgical process called normal grain growth the statistical theory of which can be found elsewhere. Its main characteristics can briefly be presented in the following way, see \cite{pande} and refs. therein: \\
(i) The polycrystalline evolution process takes place under a certain constraint concerning the constancy of the total system volume; 
(ii) It relies on evolution of grains constituting a specimen in such a way that the mean radius of the grains has to increase (the curvature(s) to diminish, automatically) while their total number has to decrease;  
(iii) An observed physical tendency while going over sufficient number of time spanes (decades) of the process is slowing down of the grain surface free energy in such a way that a possibly low surface free energy matter  agglomerate is eventually obtained. 

Put the above another way, let us follow C.~S. Smith \cite{smith} who  argued that normal grain growth results from a certain subtle, and recently re--examined \cite{mullins,pande}, interplay between topological {\it viz} geometrical (polygons--containing) or space--filling but random requirements  as well as those of thermodynamic nature, preferentially due to surface tension agitation as well as elasticity--to--rigidity (bending) effects on the grain boundaries when the system really "fights" to attain a local thermodynamic equilibrium under a set complex physico--chemical circumstances it meets on its way \cite{novikov}. 

Therefore, the role of curvature seems to us to be absolutely essential \cite{peczak}. Because,
however, of differences in the size of the main constituting entity (grain; cluster) within the
classes of mesoscopic systems we want to examine, we see that for some bigger (coarse--grained, less elastic micrometer--scale) systems, as those metallurgical, the conception of  the mean curvature would likely suffice to describe the system satisfactorily whereas for description of fine--grained assemblies (more elastic and less brittle nanosystems) the role of the so--called Gaussian curvature must be underscored 
\cite{agluczka}. 
All the rationale we have just developed above finds its right as well as well--established place in some thermodynamic observations first published by Tolman \cite{tolman} who noticed that in certain fluid--like systems of suitable (appreciably small) size a size--dependent correction to surface tension improves its description. This led then to extensive studies of surface tension size--dependent effects in versatile Van--der--Waals as well as other systems \cite{bedo,tolman}.  

Since the whole statistical--mechanical kinetic description of the grain growth of normal type (being a Fokker--Planck--Kolmogorov (FPK)--type description\footnote{In some literature called also a Smoluchowski or generalized diffusion approach.}), or with uniform grain boundaries \cite{mullins} can be found elsewhere, we encourage a reader to consult \cite{niemiec,agad1,agad2}. An important thing to mention here is that the corresponding grainy matter flux has been constructed in a phenomenological way therein \cite{agad1}, and that in it a standard--in--form diffusion (1st Fick's law) term has been completed by a drift curvature - (capillarity) dependent term, just for reflecting properly the essential mechanism governing  the presumably ordered agglomeration on many nuclei \cite{novikov,mullins,pande}. 

Very recently, another more improved way of constructing the matter flux has been proposed. It is
based on dealing with a system undergoing normal grain growth as open thermodynamic system, and on
applying the principles of nonequilibrium thermodynamics in the space of the grain sizes. (This
proposal can also be extended on the time domain.) It immediately leads to express the entropy
production, by the afore mentioned flux as well as by the adequate chemical
potential gradient. The gradient, in turn, is to be defined by the chemical activity, $a$, being not equal to one for non--ideal systems, and by the physical potential, designated by $\Phi $, that can be selected in a suitable form, conforming to physical circumstances we want to address, {\it cf.} micro-- or nanometer--scale evolutions. The procedure just outlined can be found elsewhere \cite{rubi1,rubi2,rubi3}. 

The article is structured as follows. In Section 2, we concisely describe a derivation of the flux of grainy matter, $J$, based on the nonequilibrium thermodynamics formalism, introduced very recently \cite{rubigad}, briefly referring in the section's beginning to the phenomenological construction proposed before \cite{agad1,agad2}. In the next two sections, Secs. 3, 4, respectively, we discuss two important physical cases leading to application of logarithmic as well as of Coulomb (or, gravitational) potentials, what refers to both afore mentioned micro-- as well as nano--world evolutions within the agglomerating system under study \cite{agluczka}. A final address and perspective (Sec. 5) complete our study on the entropic barriers in matter agglomeration. 

\section{Nonequilibrium thermodynamics derivation of a normal grain growth model}

A procedure followed before \cite{agad1,agad2,agluczka} relied on proposing the flux of grainy matter yielding the agglomerate's output in a phenomenological form \cite{rubigad}
\begin{eqnarray}
\label{J}
J(v, t) = - \sigma v^{\alpha -1} f(v,t)
 - D v^{\alpha} {\partial \over \partial v}  f(v,t), 
\end{eqnarray}
which implies essentially a decomposition into two parts. Here $\sigma $ and $D$ are surface tension as well as diffusion reference\footnote{The meaning of the word 'reference' is the following here: $\sigma $, being a positive constant, corresponds to the surface tension of a flat interface (in consequence, the reciprocal of the flat interface radius, $v^{\alpha -1}$, gets zero, causing the drift term in Eq. (\ref{J}) to vanish), whereas the constant $D$ is the diffusion coefficient taken from a standard--diffusion Einstein--like realization of the process in $d=1$, where the term ${v^\alpha} = 1$, see Eqs. (\ref{J}) and (\ref{alfa}).} 
constants, respectively, and in general 
$\sigma \ne D$  is true; the internal state variable $v$ and an independent variable $t$ represent the volume of an individual grain as well as the time, respectively. $\alpha $ stands for a dimensionality $d$ (where $d=1,2,3, ...$) dependent exponent \cite{agad1}
\begin{eqnarray}   \label{alfa} 
\alpha = {{d-1}\over d}, 
\end{eqnarray}
{\it i.e.} it is presented in a fractional form, namely, by providing a ratio of the subdimension $d-1$ and the space dimension $d$. (This can be thereafter referred to as a surface--to--volume exponent; notice that when $\sigma = \alpha $ holds, one arrives at a standard description of normal grain growth by Mulheran and Harding, whereas for $d=1$, one gets $\alpha  = 0$ ($\sigma = 0$), and a purely diffusive description of possibly ordered agglomeration by Louat, {\it cf.} \cite{niemiec,agad1,agad2,rubigad}, and refs. therein.)  The first part of the right-hand side (r.h.s.) of Eq. (\ref{J}) is a drift term. The second part of the r.h.s. of Eq. (\ref{J}) stands for a diffusion term. The former is a capillary term, proportional to the (mean) grain curvature, and much involved in the Laplace--Kelvin--Young law, so much explored for crystallites--containing systems \cite{novikov,agluczka}. 
The latter is simply the first Fick's law, and it underscores a proportionality of the flux to the grain surface magnitude. Both the terms incorporated in (\ref{J}) are based on phenomenological laws, so is, without doubt, the construction of the mixed convection-diffusion local flux (\ref{J}), supplemented here by (\ref{alfa}) being of geometrical meaning. Bear in mind here that via a relation $v\propto R^d$ ($R$ - a grain radius) in the drift term one really provides $v^{\alpha -1} \propto R^{-1}$ (see, Eq. (\ref{alfa})), which  means, that a curvature term can easily be revealed this way. In the diffusion term, in turn, one clearly gets a surface term $R^{d-1}$ since 
$v^{\alpha} \propto R^{d-1}$ when $v\propto R^d$ is applied again.  

After stating explicitely the flux, we have to apply it to the local continuity  equation   
\begin{eqnarray}
\label{Eq}
\frac{\partial}{\partial t} f(v,t)   
+ {\partial \over \partial v} J(v, t) = 0, 
\end{eqnarray}
where $v$ is the volume of a grain, $f(v,t)$ is the distribution function of the grains
 at time $t$ (having a meaning of the number density, but being unnormalizable \cite{niemiec,agad1}), that means, $f(v,t)dv$ is a relative number  of grains of size in the volume interval $[v,v+dv]$. 

The overall conceptual and mathematical construction is to be completed by suitable initial and boundary conditions. 
A general observation is, however, that the process in question does not depend upon any prescribed initial condition, {\it cf.} \cite{niemiec}. After a sufficient time interval being overcome it completely forgets its initial state \cite{pande}. 

This is not the case of the boundary conditions (BCs), however. The process visibly depends upon BCs prescribed. 
A choice that remains as most explored is the choice of zero Dirichlet BCs, namely 
$f(v=0,t) = f(v= \infty,t) =  0$.   
(Realize that the BCs--proposal may not work in a semi--infinite phase space, $v\in[0,\infty ]$; another use of BCs can lead to  some inconsistency, because the volume of an   
individual grain can be large and can even be larger than 
 the overall volume of the system, {\it cf.} discussion in \cite{agluczka,niemiec,rubigad}.) 

There is some physics staying behind the BCs: The grains of zero as well as of infinite sizes have zero account for the grain growth. This physical constraint is sometimes called a normality condition, and the process is supposed to be normal when the above is true. Otherwise, one may name the agglomeration abnormal \cite{novikov}. 

Suppose that we are still interested in a construction procedure of the flux $J(v,t)$, hopefully arriving at a form reminiscent of that of Eq. (\ref{J}).  It is worth doing, first,  because the phenomenological way gives us no full satisfaction and/or control, and second, that we wish to have a more sophisticated procedure staying behind the derivation, with possible open ways for application as well as more versatile examination of the polynuclear agglomerations. So, we realize that another non--phenomenological method will be applied therein though we will be not capable of avoiding phenomenology totally \cite{rubigad}. We will, however, be capable of arguing quite firmly and in a physically reasonable as well as quite general way on the basic mechanism of the agglomeration in the spirit of incorporating (or, having not incorporated) the suitable entropy barrier as a geometrical one \cite{rubipre}, here inevitably associated with the principal grain curvatures. Another, more explicit as well as presenting really state of the art (geometry) way has been shown in \cite{rubipre}, and via a Random Walk (RW)--analogy invoked below, can be adopted in a future task for deriving specific geometry-dependent forms of the physical potentials upon request. 

Thus, assume that the continuity equation (\ref{Eq}) is still the proper  evolution equation for the agglomerating system but the flux $J(v,t)$ remains to be specified. Assume, moreover, that there is a method of deriving it, and that the method is called mesoscopic nonequilibrium thermodynamics \cite{rubi1,rubi2}. It starts with the Gibbs equation which represents the entropy variations, namely \cite{rubi1,rubi3}
\begin{eqnarray}       \label{Gibbs}
\delta S = - {1\over T} {\int \mu (v,t) \delta f \;dv} ,
\end{eqnarray}
where $f\equiv f(v,t)$, $T$ is the temperature, and $\mu (v,t)$ is the chemical potential in $v$-space. The latter is given by (here, $\mu \equiv \mu (v,t)$ for a brief notation is taken)
\begin{eqnarray}       \label{cpot}
\mu = {k_B} T ln (a f) ,
\end{eqnarray}
where $a\equiv a(v)$ is an activity coefficient reflecting the fact that the system is not ideal; for $a=1$ it would become ideal. (Bear in mind that the $ln$--form of the chemical potential suggests the system were ideal but,  fortunately, the condition $a\ne 1$ would likely violate  the presupposed ideality.) $k_B$ is the Boltzmann constant. Next, $a$ is given in terms of a potential that for the reason of a clear differentiation against the chemical potential $\mu $ let us call the physical potential, $\Phi $. $a$ reads now \cite{rubi1,rubi2}
\begin{eqnarray}       \label{ppot}
a = e^{\Phi/{k_B T}}.
\end{eqnarray}
Providing the temporal derivative in Eq. (\ref{Gibbs}) and partially integrating (assuming, additionally, that $J\equiv J(v,t)$ vanishes at the ends of the phase space), one gets the entropy production, $\sigma _E$, in $v$-space
\begin{eqnarray}       \label{entropy}
{\sigma _E} = - {1\over T} J {{\partial \mu}\over  {\partial v}} 
\end{eqnarray}
from which we may easily infer the expression for the grainy flux 
\begin{eqnarray}       \label{current}
J = - {1\over T} L(v) {{\partial \mu}\over  {\partial v}}. 
\end{eqnarray}
(Here, we have assumed that the process is local in $v$; one would also consider, if necessary, a non--local case by $J(v) = - {1\over T} \int {dv'} L(v,v') {{\partial \mu}\over  {\partial {v'}}}$, {\it cf.}  \cite{rubi1,rubi2}.) 
Combining Eqs. (\ref{current}), (\ref{cpot}) and (\ref{ppot}) one gets 
\begin{eqnarray}       \label{current1}
J = - {1\over {T f}} L(v) \bigg[{{k_B T}{\partial f\over  \partial v}+ f {\partial \Phi\over  \partial v}}\bigg]. 
\end{eqnarray}
Now, let us define the mobility $b(v)$ as $b(v) = {1\over {T f}} L(v) = {D\over {k_B T}} {v^\alpha }$, where $D$ and $\alpha $ are stated as above. 
 
The derived flux $J$ is given by 
\begin{eqnarray}
\label{JJ}
J = - D v^{\alpha} {\partial \over \partial v}  f  - b(v) f {\partial \over \partial v}  \Phi .
\end{eqnarray}
The obtained expression, {\it viz}  Eq. (\ref{JJ}), looks quite general, probably inspite of the power--law form (grain--volume correlations) assumed in the Onsager coefficient $L(v)$. The main physics to anticipate now is to propose a valuable expresion for the physical potential $\Phi = \Phi (v)$. This is, however, a matter of offering a principal agglomeration mechanism for the system(s) to evolve. A proposal for a micrometer-- \cite{kurz} as well as nanometer-- \cite{rubi3} sized evolutions, respectively, will be provided in the two subsequent sections. 

\section{Matter agglomerations in micrometer--world}

Our first proposal is to make use of that there is a possibility of including an above announced geometrical barrier when a non--power (here, logarithmic) form of the potential $\Phi = \Phi (v)$ is presumed \cite{rubigad}. We see that the way of doing it is worth going, see Introduction. This is generally because of the well--known Tolman correction in curvature performed for the surface tension of a "droplet" of radius $R$, namely $\sigma \simeq 1-({\delta _T}/R)$, where $\delta _T$ stands for the Tolman length. Such a correction procedure, going to predict a decrease of the surface (line) tension for sufficiently small objects,  leads also to an inclusion of (twice) the mean curvature (the entire  thermodynamic context assumes the Gibbs--Thompson or capillary length,  $\delta _{GT}$, to be a  characteristic length of surface or line tension shrinking actions) of a curvature--driven grain growth and phase separation mechanism \cite{hillert} whereas the so--called second correction, just  related with the Tolman length,  $\delta _T$ \cite{tolman,bedo}, and termed the Gaussian curvature, reveals certain elastic (towards elasticity/rigidity, and the bending effect) properties of an agglomerate's grain surface in a time instant chosen, and when going over its evolution stages, in which such characteristics can vary. 

Thus, the proposed non--power form of the physical potential is of a logarithmic form, namely 
\begin{eqnarray}       \label{fi-npower}
\Phi= {\Phi _o} {ln(v/{v_o})},    
\end{eqnarray}
where $v_o >0$ stands for the initial grain volume. Interestingly, exactly this form suits very well to what we want to get. The entire rationale we try to reveal thoroughly, mostly below, is that we wish to have in the drift term, present in either Eq. (\ref{J}) or Eq. (\ref{JJ}), a factor of $v^{\alpha -1}$ which is explicitely a clear signature of the curvature, being a reciprocal of the grain radius $R$, see Sec. 2. 
Thus, after simple algebra 
${{\partial \over \partial v}  \Phi } = {{{\Phi _o}}{1\over v}}$, we are able to recover the desired form of the drift term in flux (\ref{J}), and this way both the fluxes, (\ref{J}) and (\ref{JJ}) take firmly on the same mathematical form, presumed that $\alpha $ is that dimensionality--dependent fractional number taken from (\ref{alfa}), called above the surface--to--volume exponent. To have a really entropic barrier an additional demand on explicit dependence of the prefactor $\Phi _o$ upon the thermal energy ${k_B}T$ is needed, as applied for certain (star) polymeric complex solutions, for example \cite{waclaw}, carrying however less about that their logarithmic part of the potential is distance--dependent, whereas ours explicitely not, {\it cf.} Eq. (\ref{fi-npower}). (We may have some excuse here by realizing again that $v\propto R^d$, where $d$ as above, and the radius $R$ can likely better mimic a distance, or a position measure than $v$ does.) 

Another issue that can be addressed seems more fundamental. It concerns our analogy between the RW in a position space and the walk along the grain size axis \cite{agluczka,agad2}, see also beginning of Sec. 2. There are some experimental evidences \cite{kurtz} as well as theoretical predictions \cite{pande,agluczka} (look at refs. therein) that the walk in grain--size space would belong to a broad class of geometric Brownian motion (a Wiener processs), the probability distribution of which is a logarithmic Gaussian \cite{agluczka}. Thus, instead of $v$ in the mathematical form of the Gaussian distribution a $ln(v)$ must appear as argument. Therefore such a potential seems legitimate too. (A certain analogy between the mass--exchange action between neighboring grains \cite{mullins} of a polycrystal, in which one of them is active, or 'waken', in transferring matter from its neighbor's volume via the grain boundary while the remaining one 'sleeps' when doing it, and the RW in a $ln$--potential, modeling sleep--wake dynamics during sleep \cite{stanley}.)

Accepting that the key problem, because of the RW-analogy in both the spaces mentioned,
does not totally rely on  whether we have $v$ or $R$ in potential function argument, {\it
i.e.} a distance measure in (\ref{fi-npower}), we may at least for a two-dimensional case,
invoke another experimental study with a logarithmic potential. The study is on a confined
mesoscopic system in which the so--called Wigner islands emerge. These are charged balls
interacting via an electrostatic potential \cite{wigner}. They represent the vortices in
type-I superconducting systems. It turns out that just the logarithmic interaction
potential causes maximum compaction of the charged spheres of milimeter, {\it viz}
macroscopic size, usually moving and agglomerating on a conductor plane \cite{wigner}. This
is thus probably no surprise that the $ln$--potential (\ref{fi-npower}) but not a power--like, for example, gives more chances for the system to advance into a constant total volume regime as the system growing under normal conditions \cite {agluczka,agad1} does. This is, as is stated in Sec. 2, the case when the governing  mechanism based on (mean) curvature driving prevails, and in which a surface (line) tension effect cannot be avoided. 

In \cite{rubi1} it was shown that diffusion as well as mobility functions are very related to each other. They are also related with the Onsager coefficient $L(v)$ via the grain (cluster) distribution function $f$, or also via an inclusion of the temperature $T$ as is in the case of mobility, and does not apply for the diffusion coefficient. 

The diffusion coefficient, in its full form denoted by $D(v,t)$, is generally assumed to be defined by the Green--Kubo correlation formula, taken from the fluctuation--dissipation theorem, in which, as the integrands, random parts, $J^r (v,t)$, of the flux (\ref{JJ}) of grainy matter are involved. It is given by  
$D(v,t) = {{k_B}\over f} L(v) = D {v^\alpha }$, 
which means that algebraic correlations are assumed in the correlator based on the $J^r (v,t)$ \cite{rubi1,rubi3}. (Some correlation in time domain are plausible too, mostly for describing agglomerations in soft--matter reactive systems {\it viz} biomembranes here \cite{jamnik}.)

Since it has been argued \cite{mullins} (see, literature therein), and experimentally proved \cite{kurz}, that the normal grain growth processes are self--similar in many respects and when based on quite many versatile measures of self--similarity both in space (grain size) and time, it would be useful to test this observation within the framework of our modeling. It is done in the subsequent section by proposing another potential $\Phi = \Phi (v)$ being generally of the Lennard--Jones (LJ) form.  

\section{Matter agglomerations in nanometer--world} 

Our second proposal on how to perform a model agglomeration is to assume, inspite of the algebraic (self--similar--in--time) correlations $<J^r (v,t) J^r (v',t')>$ of the flux $J^r (v,t)$ \cite{rubi1}, that also the 
physical potential is of a power form in $v$ 
\begin{eqnarray}       \label{fi-power}
\Phi=  {{\Phi _1}\over {v^{\epsilon _1}}} + {{\Phi _2}\over {v^{\epsilon _2}}}, 
\end{eqnarray}
where $\Phi _i$  ($i=1,2$) are constants (they may depend upon temperature, and as suggests the construction of the LJ--potential, they must be of opposite signs so that we choose here $\Phi _1 <0$, {\it i.e.} the attractive part of the LJ--potential has to be taken with minus sign, so that the Van der Waals--like attraction for preserving agglomeration is desired), and $\epsilon _i$ ($i=1,2$) are positive exponents. 

We have now to perform partial differentiation,  ${{\partial \over \partial v}  \Phi }$, and put the resulting expression into Eq. (\ref{JJ}), which gives 
\begin{eqnarray}
\label{JJp}
J = - D v^{\alpha} {\partial \over \partial v}  f  - \sigma _1 {v^{\chi _1}} f  - \sigma _2  {v^{\chi _2}} f ,
\end{eqnarray}
where: ${\chi _1} = \alpha - \epsilon _1 - 1$, ${\sigma _1} = - {\Phi _1} \epsilon _1
{{\eta}^{-1}}$; ${\chi _2} = \alpha - \epsilon _2 - 1$ and ${\sigma _2} = - {\Phi _2} \epsilon _2
{{\eta}^{-1}} $; here ${{\eta}^{-1}} = {D\over {k_B T}}$, is the inverse of a viscosity. 
This way, we get the grainy matter flux (\ref{J}), or strictly speaking (\ref{JJ}), in a very satisfactory form. A direct comparison with (\ref{J}) from Sec. 2 provokes to let $\epsilon _i\to 0$ (${\chi _i} \to \alpha - 1$) which yields $\Phi \to const$ (very weak forces as the Van der Waals normally are!). Thus, the flux (\ref{JJp}) is really in the form we await very much except that the very fact that both the surface tension reference constants $\sigma _i$  ($i=1,2$) unfortunately tend to zero, {\it cf.} the formulae below Eq. (\ref{JJp}). In this way, the drift term is completely removed, and we get no drift effect on the agglomeration. Thus, an entropic barrier due to curvature cannot be proposed based on such an oversimplified argumentation. (Notice, by the way, that in our geometry--dependent model, {\it cf.}  \cite{agad2,jamnik} and refs. therein, the prefactor appearing in the drift term of Eq. (\ref{J}) is a kind of curvature--including over--constraint always present in our thermodynamic model, recall that $v^{\alpha -1} \propto R^{-1}$, see again Eq. (\ref{J}) as well as Eq. (\ref{alfa}).)

Let us try, however, another way to put things forward. Let us take that there exists a mutual dependence between $\chi$--s. It is likely to be proved while looking at standard forms of LJ-- (or, $6-12$--potential) as well as LJ--like (for instance, $10-12$--potential, used in modeling some presence of $H$--bonds in protein assemblies, see \cite{cieplak}, and refs. therein) potentials. We see for them a simple mutual dependence between $\epsilon$--s, for instance $\epsilon _2 = 2\epsilon _1$ for the $6-12$--potential. Or, $\epsilon _2 = {(6/5)} \epsilon _1$ just for the $10-12$--potential. 
But exploration of such a simple algebraic "correlation" makes no sense since, as mentioned above, always $\sigma $--s will go to zero. In other words, there is no reason to get on the problem when staring from the potential's form directly. 

It is, however, worth exploring here a sort of mutual dependency between $\chi$--s, that means, when starting from the level of the flux (\ref{JJp}). It can be proposed as a systematic but very simple algebraic procedure, and can be done as follows. 
Namely, let us first recall that both $\chi$--s are of the form 
\begin{eqnarray}
\label{chis}
{\chi _i} = {(\alpha - 1) - \epsilon _i }, \qquad i = 1,2 ,
\end{eqnarray}
but notice that they cannot equal one another. In consequence, let us assume a simple linear relation 
\begin{eqnarray}
\label{chiq}
{\chi _2} =  {q_\chi} {\chi _1},
\end{eqnarray}
where ${q_\chi}\ne 1$, and clearly ${q_\chi}\ne 0$. After simple algebra, by making use of the equality (\ref{chiq}) as well as applying (\ref{chis}), one gets 
\begin{eqnarray}
\label{eps12}
{\epsilon _1} =  ({{\epsilon _2}/{q_\chi }}) + {(1-{q_\chi}^{-1})(\alpha -1)} 
\end{eqnarray}
or, equivalently, one can provide ${\epsilon _2} =  {{\epsilon _1} {q_\chi }} + {(1 - {q_\chi})(\alpha -1)}$ but we choose the case of $\epsilon _1$  for further consideration.

By letting $\epsilon _2\to 0$ we wash out completely the term with $\sigma _2$ in Eq. (\ref{JJp}), but we still have a nonzero $\epsilon _1$ that reads in such a limiting case as follows
\begin{eqnarray}
\label{eps12}
{\epsilon _1} =  {(1-{q_\chi}^{-1})(\alpha -1)}. 
\end{eqnarray}
This exponent for a possible choice of ${q_\chi} = {1/2}$ (${q_\chi}$ carries this way some signature of the second curvature correction, by the way) gives in Eq. (\ref{JJp}) a term which looks like 
$v^{2(\alpha -1)}$, what again via $v^{2 (\alpha -1)} \propto R^{-2}$ results in appearence of a second curvature (Gaussian) term, being proportional to $1/R^2$ \cite{bedo,tolman,hweaire}. 
This term must clearly be of importance for sufficiently small grains--containing systems, and we argue that it may happen in the nanometer scale \cite{novikov}.  Crudely speaking, it looks like the evolution in micrometer scale goes on more likely in a logarithmic fashion ($ln$--potential) than algebraically. (The size of the basic entity constituting an agglomerate is not very much emphasized in micrometer scale.) The latter, in turn, is much pronounced in nanoscale because of apparently small size of the basic constituting entity: In this physical scale, however, a LJ--like potential is very likely to be in favor, {\it cf.} the above argumentation.  

In nanoscale, {\it cf.} Eq. (\ref{fi-power}), the form of $\Phi $ is the following 
\begin{eqnarray}       \label{Coulv}
\Phi (v) =  {\Phi _2} + {{\Phi _1}\over {v^{- (\alpha - 1)}}}, 
\end{eqnarray}
what in terms of the distance (or, linear size) $R$ results in
\begin{eqnarray}       \label{CoulR}
\Phi (R) =  {\Phi _2} + {{\Phi _{11}}\over {R}},\quad \Phi _{11} < 0,\quad \Phi _2 > 0  ,
\end{eqnarray}
and $\Phi _{11}$ is a constant closely related to $\Phi _1 < 0$. 

For a microscale description, but given in terms of $R$, one provides the 
potential $\Phi (R)$ as 
\begin{eqnarray}       \label{fi-npowerR}
\Phi (R) = {\Phi _{oo}} {ln(R/{R_o})},    
\end{eqnarray}
where $\Phi _{oo}$ is a constant staying again in close relation to $\Phi _o$, whereas $R_o$ looks like an initial grain radius. Certainly, Eq. (\ref{fi-npower}) is a $v$--containing counterpart of the above, see Sec. 3. 

Of utmost importance appears to be the understanding of how does the 
grain boundary (cluster peripherial area), that is going to mimic an interface 
(here, between contiguous grains, or clusters, of possibly different internal structural architecture) curve, and upon which physical circumstances does it really go? There are two possible thermodynamic routes proposed in such a situation, namely \cite{bedo2}: \\
(i) the one, based on how does an equilibrium interface curve due to a chemical gradient change just on the grain boundary interface, \\
(ii) another one, that undergoes nonequilibrium thermal fluctuations. \\
Both the routes mentioned above seem to be plausible within the presented agglomeration context, though the scenario drawn must always depend solely upon concrete physical mechanism applied \cite{shvind,novikov,mullins}. Though the route (i) cannot be discarded {\it {\'a} priori} for our agglomeration context as a whole, it looks more safely applicable to microagglomeration driven by non--LJ potentials and forces as in some metallic as well as thin--film ceramic \cite{czekaj} systems, {\it e.g.} in relaxor multicomponent materials \cite{skulski}. The route (ii), in turn, is readily supported by the LJ--force--and --potential context as in soft--matter systems \cite{novikov,cieplak,earnshow}, {\it e.g.} in micellar droplets and biomembranes for which both theoretical (Kirkwood--Buff approach) \cite{napior} as well acurate numerical \cite{stecki} evaluations of the rigidity constant, associated with the Gaussian curvature effects on a membrane surface, predict its value to be negative, and the interactions have been assumed to be of LJ--form \cite{bedo2}, so that, according to our rationale, we have to be placed now within the nanoagglomeration context. 
This is, thus, another evidence that the Gaussian curvature is related with the LJ--context, and {\it vice versa}. The route (ii) suits very well to highly fluctuating ordered agglomerations in biopolymeric complex soft--polycrystals \cite{ag}, for which the surfaces of the grain boundaries, denoted by $s_{d=3}$, fluctuate with the time $t$ as $s_{d=3} \propto {t^2}$ for $t>>1$ again, {\it cf.} \cite{ag} and the discussion between formulae (14) and (15) therein.

Finishing this section, it is worth mentioning that both the forms derived above in terms of a distance measure $R$, Eqs. (\ref{CoulR}) and (\ref{fi-npowerR}) respectively, are physical--potential forms well--known, {\it e.g.} in electrostatics: The former represents a far--distance electrostatic potential, emerging from an electric point charge, whereas the latter is expected for extremely long cylindrical (presumably, torus--like) capacitors in which internal cylinder is charged oppositely (for example, negatively) to its external cylindrical (positive) counterpart. A qualitative observation can be offered, namely, that here a tube--like structure of the capacitor recalls somehow the capillarity effect, which prevails in the microworld but would not be a decisive phenomenon in its nano--counterpart. 
>From the textbooks on electrostatics it can be learned that both the prefactors present in the
inverse--power term of (\ref{CoulR}) as well as in (\ref{fi-npowerR}) should, in a material
medium, contain the dielectric constants of the medium they are made of. It belongs, in turn, to
a standard knowledge on the principles of biomolecular physics, that, if the dielectric constant
enters into a description of interactions, like in ligand--protein or other bio--complexes, or
agglomerates of them, there is a good opportunity to advocate for an explanation that the entropy
change (but not its enthalpic counterpart) might have the foremost influence on the agglomeration
process. Therefore, we are privileged to call, even without going into details, the entropic barriers,
being represented by the curvatures, as potentials, causing the entropy change in the
agglomerating system under study. For a typical interacting biomolecular system the entropy
contribution under normal temperature conditions around $T\simeq {298 K}$ is usually 
about $3-4$ times
larger than the enthalpy contribution can be, though they can be of opposite signs, with a
preference about a positive value for the entropic term, however.
Certainly, a formal analogy may also be extended for the gravitational objects driven
by the corresponding central forces that are based on the potentials (\ref{Coulv}) and
(\ref{fi-npower}), respectively. As for the logarithmic potential, one may also found an appealing analogy in biomemebrane science between a system consisting, {\it e.g.} of a lyposome (vesicle) with a small colloid attached, having a common logarithmic contour (catenoid) that would resemble an equipotential surface of the system mentioned, {\it cf.} \cite{helfrich}, and refs. therein. Such a scenario is reminiscent of our grains--containing agglomeration in which, on a microscopic level, a small grain is being attached to its bigger neighbor, being absorbed thereafter, preferentially because of the action of capillarity \cite{shvind,mullins,pande}. The two curvatures mentioned constantly throughout the paper are of importance in both the situations described \cite{helfrich,pande,mullins,niemiec,peczak}. 

\section{Final address and perspective}

\indent  The final address, and a perspective towards how to possibly deal with it further, can be split up in the following: 
\begin{itemize}
\item 
To sum up: We have applied mesoscopic nonequilibrium thermodynamics \cite{rubigad} in order to improve a phenomenological $d$--dependent construction \cite{agad1,agad2} of the grainy matter flux given finally in Sec. 2 by Eq. (\ref{JJ}), and proposed to describe an ordered agglomeration, termed the normal grain growth \cite{mullins}. 
\item 
When offering such an improvement we have realized that we have either to calculate explicitely the form of the correlations of the random part of grainy flux \cite{rubi2,rubi3} by relying on a physical mechanism, proposed for such a purpose, or to assume a (physically natural) geometrical constraint by keeping the surface--to--volume exponent effective, see Eq. (\ref{alfa}). It enables then to realize that the (mean) curvature has to be a principal geometrical constraint, {\it cf.} Eqs. (\ref{J}) and (\ref{JJ}).
\item It has remained to note that in "hard" condensed--matter systems \cite{mullins,pande}, mostly, the curvature, or a simple reciprocal of the grain radius, would sufficiently support the description of the system evolution. In soft condensed--matter agglomerates, in turn, operating most effectively in na\-no\-sca\-le (see, Fig. 1), the Gaussian curvature, being a reciprocal of $R$ squared, as well as the Tolman correction to surface (line) tension, emphasized above, are just the case \cite{peczak,shvind}. (By emphasizing the role of both the curvatures a kind of crossover between the nano-- as well as microscales can be addressed as well.) It would then be a serious candidate for supporting the overall evolution strategy already described. 
\item The forms of the proposed physical potentials $\Phi $ are then offered twofold: Either in a logarithmic form (\ref{fi-npower}) or as a sum of two power--like forms (\ref{fi-power}). Though a choice of the potentials known in physics is pretty vast \cite{stauffer}, we have exactly picked up those two forms because they not only follow well the relatively simple mathematics staying behind the model \cite{rubigad,niemiec} but above all they suit extremely well to the offered physical scenarios of agglomeration; in this moment, let us state clearly that, perhaps, except for entering the region of very small inverse Debye length values (high temperature limit or small co-- and counter--ion concentrations),  we see now no direct chances for applying here readily the frequently mentioned {\it DLVO} physical potential (screened Coulomb) so well applicable to most of colloid as well as surfactants--containing systems \cite{hansen,stauffer}. This is maybe therefore that it describes mostly repulsion effects but for preserving agglomeration we undoubtedly need attraction too \cite{hansen}. (By the way, if we assume $\Phi $ to be given by a single power form we will be unable to recover the Gaussian curvature mechanism at all \cite{rubigad}.)
\item The normal grain growth model \cite{mullins,pande,hillert} is a model in which agglomeration is being realized in a random fashion and under a space--filling constraint \cite{agad2}. It is well--known \cite{agad1,niemiec} that for it the kinetic dependence of a mean grain radius is given by $R\sim t^\nu$ ($t>>1$), where $\nu = 1/(d+1)$, {\it i.e.} the exponent is given by a reciprocal of the superdimension $d+1$. Interestingly, the model with a surface--to--volume exponent $\alpha$ assumed, which clearly invokes colloidal agglomerations \cite{earnshow}, see Eq. (\ref{alfa}), yields an asymptotic kinetic law constituted by the superdimension $d+1$. But it seems natural: the $d+1$--account is just a proper description of nearest neighborhood for a grain in $d$--dimensions  \cite{agad2}; the whole matter agglomeration strategy is sometimes termed in textbooks the random close--packing. Therefore, such a kinetic law seems legitimate as well \cite{niemiec,agad1}. 
\item The above is going to be a bit changed for the nanoworld for which a kinetic law may change,
presumably as $R\sim t^\omega$ ($t>>1$), where $\omega = 1/(d+2)$ could be predicted here. Thus,
the evolution goes slower. There is an evidence that such a slowing down of the grain--growth
rate may emerge, for instance in some specially fabricated (cryomilled; then kept in about
$2/3$--regime of melting temperature) nanopowders, and mostly due to pinning effects, accompanied
by precipitation sub--effects at grain boundaries, presumably giving room for Gaussian--curvature
assisting, or bending, phenomena to enter \cite{riso}. Moreover, both the models, with the mean as well as the Gaussian curvatures, are quite reminiscent of a "hydrodynamic" pairwise coalescence (ripening; clustering) model proposed in the early seventies by Binder and Stauffer, in which however, the  diffusion process is really done in a position space, being as in our model, composed of two (other) parts: a translational as well as some rotational, {\it cf.} \cite{furukawa}, and refs. therein. The kinetic radius {\it vs} time law was obtained the same \cite{furukawa}, and the cluster--interaction parameter inferred from Binder and Stauffer's  formulae, say $\iota $, obeys:  
$\iota = d-1$ (the subdimension) for the microworld evolution, whereas $\iota = d$ (the Euclidean dimension) for nanoagglomeration, both of them being exhaustively described in the present article in terms of a nonequilibrium thermodynamics model with geometric constraints. 
By the way, note that $\alpha $ of Eq. (\ref{alfa})  just combines again both $\iota$--s mentioned above. 
\item A general observation appears that anticipates the agglomeration process as being accelerated (mean curvature mechanism) or decelerated (Gaussian curvature mechanism) by presence of such entropic, kinetic--geometric, barriers; moreover, presence of the drift term in Eq. (\ref{J}), or equivalently in Eq. (\ref{JJ}), makes the agglomeration more directional {\it viz} ordered.  In all the cases mentioned the role of surface tension is to be emphasized, but its possible neglect, characteristic of loosely agglomerating systems \cite{lutz} (ripening; cluster--cluster aggregation \cite{mullins,novikov,earnshow}), leads also to the same $d+1$--account in the growth law because in late time regimes the role of capillarity, according to the frequently addressed Laplace--Kelvin--Young law, is not very much underlined, and therefore the asymptotics of both the systems, with and without visible capillarity influence, look the same \cite{niemiec,lutz,hillert}. 
\item Some possible future task remains to solve formally the LJ--potential influenced system (with the Gaussian curvature mechanism, including $v^{2 (\alpha -1)}$--term) 
\begin{eqnarray}
\label{Eqn}
\frac{\partial}{\partial t} f(v,t)   
+ {\partial \over \partial v} J_G (v, t) = 0, \nonumber
\end{eqnarray}
where 
\begin{eqnarray}
\label{Jn}
J_G (v, t) = - \sigma v^{2 (\alpha -1)} f(v,t)
 - D v^{\alpha} {\partial \over \partial v}  f(v,t), \nonumber 
\end{eqnarray}
with appropriate initial as well as boundary conditions as it was already done for its
$ln$--potential influenced counterpart \cite{niemiec,agad1,lutz,agad2,rubigad}. (But it is
left for another study, preferentially for $d=3$, {\it i.e.}, $\alpha={2/3}$, which is
physically a more transparent as well as quite common case \cite{hweaire,novikov}.) Moreover, it is possible to consider other, 
higher order (above 2nd order), accounts to curvature, what has already been done for
single--crystal growth from either supersaturated solution or undercooled melt
\cite{kessler,agluczka}. For example, for $q_\chi = {1/3}$ a third curvature correction, like
$1/R^3$, responsible for asymmetric growth \cite{kessler}, may emerge. (Higher curvature "modes" of the
growing process\footnote{In general, the distance--dependent potential $\Phi (R)$ would consequently obey: $\Phi (R) \propto {Const. + {R^{-(n-1)}}}$ for $n > 1$ and natural, where $n$-s enumerate the subsequent curvature modes.}, like ${{q_\chi}^{(n)}} = {1/n}$, where $n >3$ here (and certainly natural), would also be
responsible for some other nonlinear, and quite numerous effects, arising possibly during growth in complex
materials \cite{shvind,rubi3,novikov,peczak,kessler}. Some of the effects predicted, may advocate
for superplastic behavior of the model agglomerates just considered. It  turns even out that effect
of superplasticity can more likely be attributed to nanomaterials because they, compared to micromaterials, are composed of smaller, and possibly more curvilinear grains for which cohesion effects are rather more pronounced \cite{agluczka,grabski}.) 
\item It is also worth examining the proposed evolution equations  in terms of a fractional
FPK--dynamics \cite{metzler} because of its better sensitivity to different short, intermediate and
long time--scale events \cite{jamnik} in nucleation--and--growth phenomena, leading quite often to
formation of model biomaterials and/or biomembranes (microemulsions, too); till now it was proposed
to be solved phenomenologically by postulating the corresponding but full space-- and time
correlations in $D(v,t)$, being of a power form, {\it cf.} \cite{jamnik}. Then, some attempts of applying standard derivatives, not the fractional ones, but for a finite agglomerating system, and with fractional statistical moments of the examined growing process, have recently been started too \cite{agluczka}. 

\item An open question remains whether so described agglomerations are supposed to be named orderly
agglomerations? Clearly, on the basis of the proposed mesoscopic mechanism, involving a stochastic
variable $v \equiv v(t)$ and no formal constraints as for preserving an order within the evolving
system (the total volume, being constant, is surely not enough here), we cannot judge unambigiously whether the agglomeration, in either the microscale or in
its nano--counterpart, proceeds orderly, or not. But, at least for bioagglomerations, {\it i.e.}
the matter agglomerations involving a biomaterial, we can state so: Once an evolution way is
selected by a system more or less at random (the initial state), then this selected random formation goes via a possibly efficient
(self-organized; minimum--energy cost) way \cite{blumen}. If an unordered system is met randomly,
in turn, the evolution can then be named as unordered. (It may, however, be slightly improved in
the course of the evolution but one cannot expect drastic, here ordered, changes, without applying
equaly drastic {\it viz} decisive external stimuli.)

Last but not the least, it is clear that the proposed theory of matter agglomeration is thought of
to be an extension of the (normal) grain growth theory unquestionably suitable for pure single phase systems, in which the kinetic parabolic law is mostly expected to occur \cite{atkinson}, whereas in our approach it looks like an exception \cite{riso}, being realized exclusively for $d=1$ ($\alpha = 0$), {\it i.e.} for the one--dimensional reference state \cite{agad1}, see Sec. 2.
\end{itemize}

\section*{Acknowledgement}
{\normalsize Thanks go to Prof. K.~J. Kurzyd\l owski (Warsaw University of Technology) and Mr. Jacek Si\'odmiak (University of Technology and Agriculture, Bydgoszcz) for reading a preceding shortened version of the manuscript as well as the present manuscript, respectively.}


\newpage

\begin{figure}[H]
\begin{center}
\includegraphics[scale=1.5]{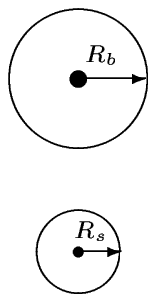}
\end{center}
\ \\

\caption{{\small
Two idealized circular grains ($d=2$): The upper grain of radius $R_b$ has both twice the mean as well as local curvatures equal to $1/{R_b}$, and represents a microagglomerate's grain, whereas the lower grain (exaggerated to be big enough for visualisation purposes), of radius $R_s$ (and the curvatures of $1/{R_s}$), is a nano--grain. The radii $R_b$ and $R_s$ are drawn in a vector representation for indicating a direction of possible local advancement of grainy matter. 
It becomes clear that the Gaussian curvature of that smaller one, ${1/{R_s}^2}$, has to be distinctly bigger than the same quantity for the micro--grain because $R_s\sim {10^{-9} m}$ is three orders of magnitude smaller than normally $R_b\sim {10^{-6} m}$ can be. (Such a size--effect, and a physical scenario associated with it, have to be taken into account while modeling agglomerates of different sizes of their basic constituting entities.) 
}}

\end{figure}

\end{document}